\documentclass[aps,prb,reprint,groupedaddress]{revtex4-1}
\usepackage{amsmath} \usepackage{amssymb} \usepackage{graphicx}
\usepackage{epsfig} \usepackage[]{csquotes} \usepackage{mathtools}
\usepackage{hyperref}

\newcommand{\D}{\text{d}}

\begin{document}

% Use the \preprint command to place your local institutional report number in
% the upper righthand corner of the title page in preprint mode.  Multiple
% \preprint commands are allowed.  Use the 'preprintnumbers' class option to
% override journal defaults to display numbers if necessary
%\preprint{}

\title{Dynamical Quantum Phase Transitions in the Kitaev Honeycomb Model}

% repeat the \author .. \affiliation  etc. as needed \email, \thanks,
% \homepage, \altaffiliation all apply to the current author. Explanatory text
% should go in the []'s, actual e-mail address or url should go in the {}'s for
% \email and \homepage.  Please use the appropriate macro foreach each type of
% information

% \affiliation command applies to all authors since the last \affiliation
% command. The \affiliation command should follow the other information
\author{Markus Schmitt}
\email[]{markus.schmitt@theorie.physik.uni-goettingen.de}
%\homepage[]{Your web page} \thanks{} \altaffiliation{}
\author{Stefan Kehrein} \affiliation{Institut f\"ur Theoretische Physik,
Georg-August-Universt\"at G\"ottingen, D-37077 G\"ottingen, Germany}

%Collaboration name if desired (requires use of superscriptaddress option in
%\documentclass). \noaffiliation is required (may also be used with the \author
%command).  \collaboration can be followed by \email, \homepage, \thanks as
%well.  \collaboration{} \noaffiliation

\date{\today}

\begin{abstract} The notion of a dynamical quantum phase transition (DQPT) was
recently introduced in [Heyl et al., Phys. Rev. Lett. \textbf{110}, 135704 (2013)] 
as the non-analytic behavior of the Loschmidt echo at critical times
% in the real time evolution of quantum systems
in the thermodynamic limit. 
In this work the quench dynamics in the ground state sector of the
two-dimensional Kitaev honeycomb model are studied regarding the occurrence of
DQPTs. For general two-dimensional systems of BCS-type it is demonstrated how
the zeros of the Loschmidt echo coalesce to areas in the thermodynamic limit,
implying that DQPTs occur as discontinuities in the second derivative. In the
Kitaev honeycomb model DQPTs appear after quenches across a phase boundary or
within the massless phase. In the 1d limit of the Kitaev honeycomb model it
becomes clear that the discontinuity in the higher derivative is intimately
related to the higher dimensionality of the non-degenerate model. Moreover,
there is a strong connection between the stationary value of the rate function of
the Loschmidt echo after long times and the occurrence of DQPTs in this model.
\end{abstract}

%\keywords{}

\maketitle

% body of paper here - Use proper section commands References should be done
% using the \cite, \ref, and \label commands
\section{Introduction} 
Recent advances in experimental techniques allow to
realize closed quantum systems with cold atomic gases in optical traps.
\cite{greiner2002,kinoshita2006} These setups are precisely controllable and
the unitary time evolution of the systems can be resolved such that the
dynamics are experimentally accessible under well-known conditions. Motivated
by the new experimental possibilities a lot of theoretical research on the
non-equilibrium dynamics of quantum systems has been conducted in the past
years. In these theoretical investigations a common protocol for driving a
system out of equilibrium is called \emph{quantum quench}. Considering a
parametrised Hamiltonian $H(\alpha)$, where the parameter typically corresponds
to some external field strength in the experimental setup, the system is initially
assumed to be in equilibrium with regard to some value $\alpha_i$ of
the parameter. Then, the parameter is suddenly \emph{quenched} to a different final
value $\alpha_f$ driving the system out of equilibrium and inducing a
non-trivial time evolution.

Studying the quench dynamics of a quantum many-body system, Heyl et al.
\cite{heyl2013} pointed out the close formal similarity of the canonical
partition function of an equilibrium system, $Z(\beta)=\text{tr}\left(e^{-\beta
H}\right)$, and the return amplitude 
\begin{align} 
\mathcal G(t)=\langle\psi_i|e^{-iHt}|\psi_i\rangle\label{eq:loschmidt_amplitude}
\end{align} 
of a time-evolved state, suggesting the possibility of critical
behavior in the time evolution in analogy to equilibrium phase transitions. It
is known that in the thermodynamic limit the zeros of a partition function
coalesce to lines in the complex temperature plane and the equilibrium phase
transition is marked by the intersection of the zero-line with the real
temperature axis.\cite{fisher1965} Heyl et al. found that in the case of the
transverse field Ising model the boundary partition function 
\begin{align}
Z(z)=\langle\psi_i|e^{-zH}|\psi_i\rangle\label{eq:bpf} 
\end{align} 
has zeros in
the complex time plane, which accordingly coalesce to lines in the
thermodynamic limit. These lines cross the real time axis after quenching the
external field across the quantum critical point inducing nonanalyticities in
the rate function of the Loschmidt echo 
\begin{align}
r(t)=&-\lim_{N\to\infty}\frac{1}{N}\ln|\langle\psi_i|e^{-iHt}|\psi_i\rangle|^2\nonumber\\
=&-\lim_{N\to\infty}\frac{1}{N}\ln\mathcal L(t) 
\end{align} 
at equidistant critical times $t_n^*$. Heyl et al. denote
this non-analytic behavior at critical times in the thermodynamic limit as a
dynamical quantum phase transition (DQPT). They showed that in experiment the
DQPT would be observable by measuring the work distribution function of a
double quench; in particular, the Loschmidt echo $\mathcal
L(t)=|\langle\psi_i|e^{-iHt}|\psi_i\rangle|^2$ is the probability of performing
no work.

These findings triggered further work aiming at a better understanding of the
phenomenon. By considering an additional integrability-breaking interaction in
the transverse field Ising chain it was demonstrated that DQPTs are not a
peculiarity specific to integrable models, but are stable against some
non-integrable perturbations.\cite{karrasch_schuricht_2013, kriel2014}
Moreover, the signature of DQPTs was found in higher-dimensional systems,
namely, in two-dimensional topological insulators\cite{vajna2014} and
effectively infinite dimensions using DMFT.\cite{canovi2014} It was observed in
various cases that DQPTs are not necessarily connected to quenching across a
quantum critical point;\cite{fagotti2013, canovi2014, vajna12014,
andraschko2014} however, there seems to be a strong connection to topological phase
transitions.\cite{vajna2014, budich_heyl_2015} Canovi et al.\cite{canovi2014} detected
coexisting solutions for so called generalized expectation values in post
quench dynamics and, therefore, they introduced the notion of a first order
dynamical phase transition. This could be a way to classify dynamical phase
transitions. Furthermore, a close connection between the analytic behavior of
$r(t)$ in the complex plane and its long time limit is conjectured.\cite{heyl_vojta_2013}

%DQPTs were to date not observed in experiments. The Loschmidt echo $\mathcal
%L(t)$, which shows the non-analytic behaviour, is not directly connected to a
%quantum mechanical observable. As mentioned above, the work density was
%suggested as an measurable quantity, which could show the signature of DQPTs.
%However, the Loschmidt echo becomes exponentially small with increasing system
%size, whereas, strictly speaking, DQPTs occur only in the thermodynamic limit.
%Measuring the signature of DQPTs in work densities in experiment will
%therefore be very challenging. 

In this work we study quench dynamics in the Kitaev honeycomb model
\cite{kitaev2006} regarding dynamical quantum phase transitions. The model
features a rich phase diagram comprising an extended gapless phase, anyonic
excitations, and topological order. Moreover, it is a rare example of a
Jordan-Wigner-solvable model in two dimensions.
\cite{chen_hu2007,chen_nussinov2008} As such it has been studied extensively
under various aspects. In this paper we restrict the discussion to the dynamics
in the ground state sector.

The dynamics of gapped two-dimensional two-band systems were already studied by
Vajna and D\'ora with focus on a connection between DQPTs and topological
phases.\cite{vajna2014} In the presence of a magnetic field the Kitaev model
acquires topological order and becomes a system of 
the same family, albeit being a spin model. 
In that case we find the behavior in accordance with their results, 
namely, DQPTs occur after quenches across the boundary between 
phases with different Chern number. However, the focus of this work lies on 
quenching between the topologically trivial phases in the absence of a magnetic field.
Similar to their results we find DQPTs as discontinuities in the second
derivative, which is inherent to the higher dimensionality of the system, and
we elaborate on the relevance of the complex zeros of the dynamical partition
function in this context. Moreover, we discuss a remarkable observation
regarding the long time behavior of the Loschmidt echo. If no DQPTs occur in
the post-quench dynamics, then, although the approached stationary state is
always an excited state, the long time limit of the Loschmidt echo is given by
the fidelity, i.e., the overlap of the initial state with the ground state of
the quenched Hamiltonian. This, however, does not hold if the dynamics exhibit
DQPTs.

The rest of this paper is organized as follows: in section
\ref{sec:kitaev_model} the way of solving the model using Jordan-Wigner
transformation is sketched and the phase diagram is introduced. Furthermore,
the expressions for the dynamical free energy is derived. In section
\ref{sec:zeros} the zeros of the dynamical partition function in the complex
time plane are treated assuming a general BCS-type Hamiltonian, yielding the
criteria for the occurrence of DQPTs and the order of the corresponding
nonanalyticity. Finally, the zeros of the partition function and the real time
evolution are studied explicitly for the Kitaev model in section
\ref{sec:kitaev_dqpt}, and two interesting limits are taken into account as well
as ramping as an alternative protocol and the quenching with an additional
magnetic field.

\section{The Kitaev honeycomb model\label{sec:kitaev_model}} 
\subsection{The model} 
\begin{figure}
\includegraphics[width=.8\columnwidth]{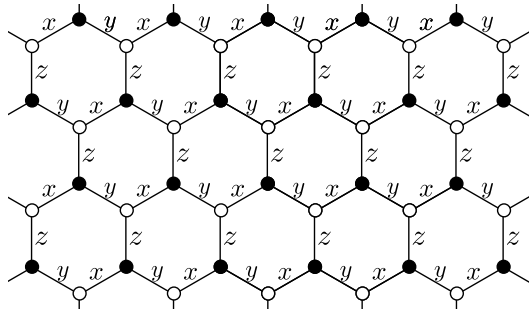} 
\caption{Lattice of the
Kitaev honeycomb model given by eq. \eqref{eq:kitaev_hamiltonian}. Spin-1/2
degrees of freedom are sitting on the vertices of a honeycomb lattice. The
nearest neighbor interaction depends on the link type ($x,y$, or
$z$).\label{fig:kitaev_lattice}} 
\end{figure}

The Kitaev honeycomb model is defined by the Hamiltonian 
\begin{align}
H(\vec J)=
-\sum_{\alpha\in\{x,y,z\}}\sum_{\alpha\text{-links}}
J_\alpha\sigma_j^\alpha\sigma_k^\alpha\
,\label{eq:kitaev_hamiltonian} 
\end{align} 
which describes a spin-1/2 system
with the spins located on the vertices (labeled by $j,k$) of a honeycomb
lattice as depicted in Fig. \ref{fig:kitaev_lattice}.\cite{kitaev2006} In this
paper we assume the lattice spacing to equal unity.  
It has been
shown\cite{chen_nussinov2008} that for the above Hamiltonian one can find a
Jordan-Wigner contour, which after identifying a conserved $Z_2$
operator\footnote{Here, we choose $\alpha_r=-1$ (cf. Ref.
\onlinecite{chen_nussinov2008}), such that the result for the spectrum agrees with
the result in Ref. \onlinecite{kitaev2006}} and switching to momentum space yields a
BCS-type Hamiltonian 
\begin{align} 
H(\vec J)=\sum_{\vec
k}&\left[\frac{\epsilon_{\vec k}(\vec J)}{2}\left(d_{\vec k}^\dagger d_{\vec
k}-d_{-\vec k}d_{-\vec k}^\dagger\right)\right.\nonumber\\
&\left.+\frac{\Delta_{\vec k}(\vec J)}{2}\left(d_{\vec k}^\dagger d_{-\vec
k}^\dagger+d_{-\vec k}d_{\vec k}\right)\right]\label{eq:jw-hamiltonian}
\end{align} 
with 
\begin{align} 
\epsilon_{\vec k}(\vec J)&=2(J_z+J_x\cos(k_x)+J_y\cos(k_y))\ ,\nonumber\\ 
\Delta_{\vec k}(\vec J)&=2(J_x\sin(k_x)+J_y\sin(k_y))\ .  
\end{align} 
This Hamiltonian can be
diagonalised by a Bogoliubov transformation 
\begin{align}
\left(\begin{matrix}a_{\vec k}^{\vec J}\\{a_{-\vec k}^{\vec
J}}^\dagger\end{matrix}\right)=\left(\begin{matrix}u_{\vec k}(\vec J) & v_{\vec
k}(\vec J)\\-v_{\vec k}(\vec J)^*&u_{\vec k}(\vec
J)^*\end{matrix}\right)\left(\begin{matrix}d_{\vec k}\\d_{-\vec
k}^\dagger\end{matrix}\right), 
\end{align} 
where 
\begin{align} u_{\vec
k}(\vec J)&=\sqrt{\frac12\left(1+\frac{\epsilon_{\vec k}(\vec J)}{E_{\vec
k}(\vec J)}\right)}\ ,\nonumber\\ 
v_{\vec k}(\vec J)&=\text{sgn}(\Delta_{\vec k}(\vec
J))\sqrt{\frac12\left(1-\frac{\epsilon_{\vec k}(\vec J)}{E_{\vec k}(\vec
J)}\right)} 
\end{align} 
(see appendix \ref{app:bogoliubov} for details).
Plugging the transformation into eq. \eqref{eq:jw-hamiltonian} yields the
diagonal Hamiltonian 
\begin{align} 
H(\vec J)=\sum_{\vec k\in K}\frac{E_{\vec k}(\vec
J)}{2}\left(a_{\vec k}^\dagger a_{\vec k}-a_{-\vec k}a_{-\vec k}^\dagger\right)
\end{align} 
with spectrum 
\begin{align} 
E_{\vec k}(\vec J)=
\sqrt{\epsilon_{\vec k}(\vec J)^2+\Delta_{\vec k}(\vec J)^2}\ .
\end{align} 
The Hamiltonian splits into a sum over $\vec k$-sectors, i.e., a sum over 
$\vec k\in K$, where $K$ is a subset of the Brillouin-zone such that 
$\forall \vec k\in K$, $-\vec k\notin K$. A possible choice is one half of the
Brillouin-zone, e.g., all $\vec k$ with $k_x>0$.\footnote{To be precise one
would have to decide how to deal with the $k_x=0$-axis; however, this will 
not play any role in the later calculations.}

\begin{figure} 
\includegraphics[width=.6\columnwidth]{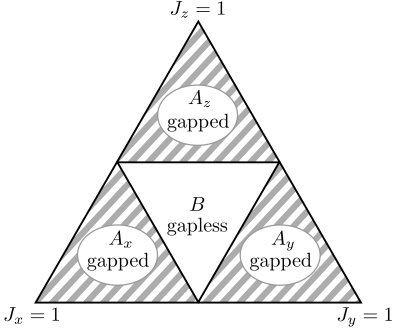}
\caption{Phase diagram of the Kitaev model in the $J_x+J_y+J_z=1$ plane, where
$J_\alpha\geq0$. The gapless phase $B$ is surrounded by three gapped phases
$A_\alpha$, where $J_\alpha>J_\beta+J_\gamma$.\label{fig:phase_diagram}}
\end{figure}

The spectrum has roots at 
\begin{align}
k_x&=\pm\arccos\left(\frac{J_y^2-J_x^2-J_z^2}{2J_xJ_z}\right),
\nonumber\\
k_y&=\mp\arccos\left(\frac{J_x^2-J_y^2-J_z^2}{2J_yJ_z}\right) 
\end{align} 
if $|J_\alpha|<|J_\beta|+|J_\gamma|$, where $(\alpha,\beta,\gamma)$ 
is any permutation of $(x,y,z)$. We will in the following only consider 
non-negative $J_\alpha$ on the $J_x+J_y+J_z=1$ plane. 
In this section the above result on the
gappedness of the spectrum corresponds to a phase diagram as depicted in Fig.
\ref{fig:phase_diagram}. The gapless phase $B$ at the center of the diagram is
surrounded by three distinct gapped phases\cite{kitaev2006} $A_x$, $A_y$ and
$A_z$, where $J_x>J_y+J_z$, $J_y>J_x+J_z$ or $J_z>J_y+J_x$, respectively.

In the presence of a magnetic field $\vec h$ the spin Hamiltonian
\eqref{eq:kitaev_hamiltonian} becomes 
\begin{align} 
H(\vec J,\vec h)=-\sum_{\alpha\in\{x,y,z\}}\left(\sum_{\alpha\text{-links}}
J_\alpha\sigma_j^\alpha\sigma_k^\alpha
+\sum_jh_\alpha\sigma_j^\alpha\right)\label{eq:ham_magn}
\end{align} 
and the additional term opens a gap also in the $B$-phase.
Moreover, the $B$-phase becomes topologically non-trivial with Chern-number
$\nu=\pm1$, whereas the $A_\alpha$ phases remain trivial with
$\nu=0$.\cite{kitaev2006} At $J_x=J_y=J_z=J$ there exists a diagonal form of
the Hamiltonian even with non-zero magnetic field\cite{kitaev2006} and the
spectrum reads 
\begin{align} 
E_{\vec k}(J,h)=\sqrt{\tilde\epsilon_{\vec
k}(J,\kappa)^2+|\tilde\Delta_{\vec k}(J,\kappa)|^2}\label{eq:magn_spec}
\end{align} 
with $\tilde\epsilon_{\vec k}(J,\kappa)=\epsilon_{\vec k}(\vec J)$,
$\vec J=(J,J,J)^T$ and 
\begin{align} 
\tilde\Delta_{\vec k}(J,\kappa)=&
\Delta_{\vec k}(\vec J)\nonumber\\
&+4i\kappa\left(\sin(k_x)-\sin(k_y)+\sin(k_y-k_x)\right)\ , 
\end{align} 
where $\kappa\sim\frac{h_xh_yh_z}{J^2}$. Through a Bogoliubov 
transformation (see appendix \ref{app:bogoliubov}) this maps to 
\begin{align}
H(J,\kappa)=\sum_{\vec k\in K}&\left[\frac{\tilde\epsilon_{\vec
k}(J,\kappa)}{2}\left(d_{\vec k}^\dagger d_{\vec k}-d_{-\vec k}d_{-\vec
k}^\dagger\right)\right.\nonumber\\ &\left.+\frac{\tilde\Delta_{\vec
k}(J,\kappa)}{2}d_{\vec k}^\dagger d_{-\vec k}^\dagger+\frac{\tilde\Delta_{\vec
k}(J,\kappa)^*}{2}d_{-\vec k}d_{\vec k}\right]\ .\label{eq:ham_magn_dirac}
\end{align} 
This case will be studied in section \ref{subsec:magnfield}.
Before, we will stick to the case without magnetic field, i.e., real valued
$\epsilon_{\vec k}$ and $\Delta_{\vec k}$.

\subsection{Post-quench dynamics} 
In order to study the dynamics in the Kitaev
honeycomb model we will consider situations where the system is initially, at
$t<0$, prepared in the ground state of $H(\vec J_0)$, i.e., $H(\vec
J_0)|\psi_i\rangle = E_\text{GS}|\psi_i\rangle$. In terms of the free fermion
degrees of freedom the initial state is the vacuum: $a_{\vec k}^{\vec
J_0}|\psi_i\rangle=a_{\vec k}^{\vec J_0}|0;\vec J_0\rangle=0$. At $t=0$ the
parameter is quenched to its final value $\vec J_1$, such that for $t>0$ the
time-evolved state is $|\psi(t)\rangle = e^{-iH(\vec J_1)t}|\psi_i\rangle$.
Making use of the Bogoliubov transformation the initial state can be expressed
in terms of the final free fermions, which diagonalise $H(\vec J_1)$:
\begin{align} 
|\psi_i\rangle=\mathcal N^{-1}{\prod_{\vec k\in K}}\left(1+B_{\vec
k}(\vec J_0,\vec J_1){a_{\vec k}^{\vec J_1}}^\dagger {a_{-\vec k}^{\vec
J_1}}^\dagger\right)|0; \vec J_1\rangle\label{eq:initial_state} 
\end{align}
Here, 
\begin{align} 
B_{\vec k}(\vec J_0,\vec J_1)&\equiv\frac{V_{\vec k}(\vec
J_0,\vec J_1)}{U_{\vec k}(\vec J_0,\vec J_1)}\nonumber\\ 
&=\frac{u_{\vec k}(\vec
J_0)v_{\vec k}(\vec J_1)-u_{\vec k}(\vec J_1)v_{\vec k}(\vec J_0)}{u_{\vec
k}(\vec J_1)u_{\vec k}(\vec J_0)+v_{\vec k}(\vec J_1)v_{\vec k}(\vec J_0)}
\end{align} 
and the normalization constant 
\begin{align} 
\mathcal N^2\equiv{\prod_{\vec k\in K}}\left(1+
B_{\vec k}(\vec J_0,\vec J_1)^2\right)\label{eq:norm} 
\end{align} 
were introduced. A more detailed
derivation is given in appendix \ref{app:bogoliubov}.

For the sake of brevity and lucidity we will in the following refrain from
dragging along the dependencies on $\vec J_0$ and $\vec J_1$ explicitly, i.e.,
identify $B_{\vec k}\equiv B_{\vec k}(\vec J_0,\vec J_1)$ and $a_{\vec k}\equiv
a_{\vec k}^{\vec J_1}$.

Using \eqref{eq:initial_state} to compute the dynamical partition function we
obtain 
\begin{align} 
Z(z)&=\langle\psi_i|e^{-zH}|\psi_i\rangle\label{eq:dpf}\\
&={\prod_{\vec k\in K}}\frac{1+B_{\vec k}^2e^{-2E_{\vec k}(\vec J_1)z}}{1+B_{\vec
k}^2} \ .\label{eq:bcs_dyn_part_func} 
\end{align} 
The dynamical partition
function has large deviation form $Z(z)\sim e^{-N f(z)}$, where $N$ is the
system size. Thus, in the thermodynamic limit only the rate function, or
dynamical free energy density, 
\begin{align}
f(z)&=-\lim_{N\to\infty}\frac{1}{N}\ln( Z(z)) 
\end{align} 
is well defined.

\section{Zeros of the partition function and critical times\label{sec:zeros}}
\subsection{General aspects} 
From the study of equilibrium phase transitions it
is known that a very insightful approach is to consider the zeros of the
partition function in the complex temperature or complex magnetization plane,
respectively.\cite{yang_lee_1952, fisher1965, bena_review_2005} In the
thermodynamic limit the zeros of the partition function coalesce to lines or
areas in the complex plane, which mark the critical points when approaching the
real temperature (magnetization) axis. Analogous reasoning has proven useful in
the study of dynamical quantum phase transitions.\cite{heyl2013, vajna2014}

In particular, an interesting analogy to electrodynamics allows to characterize the
dynamical phase transition through the density of zeros of the dynamical
partition function in the complex time plane. The starting point is the
observation that the dynamical partition function \eqref{eq:dpf} is an entire
function of $z$ and can as such, according to the Weierstrass factorization
theorem, be written as 
\begin{align}
Z(z)=e^{h(z)}\prod_{j\in J}\left(1-\frac{z}{z_j}\right)\ , 
\end{align} 
where $J$ is some discrete index set,
$z_j\in\mathbb C$ are the zeros, and $h(z)$ is an entire function
\cite{heyl2013}. With this, the dynamical free energy density reads
\begin{align}
f(z)=-\lim_{N\to\infty}\frac{1}{N}\left[h(z)+\sum_{j\in J}\ln\left(1-\frac{z}{z_j}\right)\right]\ .
\end{align} 
From this expression it becomes clear, that any non-analytic
behavior of the dynamical free energy can only occur at or in the vicinity of
the zeros of the dynamical partition function $z_j$. Since we are interested in
nonanalyticities, we will in the following ignore the contribution of $h(z)$
and only consider the singular part 
\begin{align}
f^s(z)=-\lim_{N\to\infty}\frac{1}{N}\sum_{j\in J}\ln\left(1-\frac{z}{z_j}\right)
\end{align} 
In the thermodynamic limit the sum becomes an integral over some
continuous variable $x\in X$, where $X\subseteq\mathbb R^n$ is a region 
corresponding to the previously used index set $J$, 
and the zeros become a function of this variable
$\tilde z(x)$, such that 
\begin{align} 
f^s(z)=-\int_X \D x\ln\left(1-\frac{z}{\tilde z(x)}\right)\ .  
\end{align} 
A transformation of the integration variable yields 
\begin{align} 
f^s(z)=-\int_{z(X)} \D\tilde z\rho(\tilde z)\ln\left(1-\frac{z}{\tilde z}\right)\ , 
\end{align} 
where the
Jacobian determinant $\rho(\tilde z)$ can be interpreted as the density of
zeros in the complex plane.\cite{saarloos_kurtze_1984} Moreover, setting
$\rho(z)\equiv0$ for $z\not\in z(X)$ allows to extend the integration domain to
the full complex plane. We will now discuss the real part 
\begin{align}
\phi(z)=\text{Re} \left[f^s(z)\right]=-\int_{\mathbb C}\D\tilde z \rho(\tilde
z)\ln\left|1-\frac{z}{\tilde z}\right|\ .\label{eq:phi} 
\end{align} 
We will later see that the Loschmidt echo on the real time axis is directly given by
$\phi(t)$. For $z=x+iy$ with $x,y\in\mathbb R$ $\ln|z|$ is the Green's function
of the Laplacian $\Delta_{2D}=\frac{\partial^2}{\partial
x^2}+\frac{\partial^2}{\partial y^2}$, i.e.,  
\begin{align}
\Delta_{2D}\phi(z)=-2\pi\rho(z)\ .\label{eq:laplace} 
\end{align} 
In other
words, the real part of the dynamical free energy density can be interpreted as
the electrostatic potential $\phi(z)$ produced by a charge density $\rho(z)$ in
two dimensions and the question of the behavior of the free energy at
critical points becomes the question of the behavior of the electrostatic
potential at surfaces. If the zeros form lines in the complex plane, this
allows to deduce the order of the phase transition directly from the density of
zeros at or in the vicinity of the physically relevant $z$.
\cite{bena_review_2005}

\begin{figure} 
\includegraphics[width=.6\columnwidth]{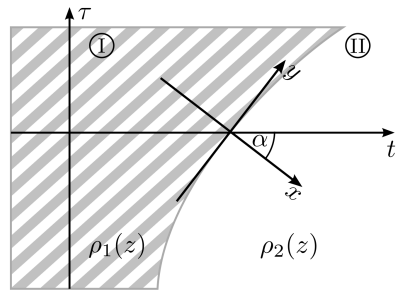}
\caption{Schematic picture of the surface separating two densities of zeros in
the complex plane and the relevant coordinate frames for determining the
behavior of $\phi(z)$ (eq. \eqref{eq:phi}) along the real time
axis.\label{fig:density_area}} 
\end{figure} 
Although the zeros can form areas
in the complex plane these areas do not cover the physical axis in the case of
equilibrium phase transitions. However, as we will see in the following
section, this is possible in the case of dynamical phase transitions. Thus,
consider the situation as depicted in Fig. \ref{fig:density_area}. The density
of zeros is given by $\rho_1(z)$ and $\rho_2(z)$ in area $I$ and $II$,
respectively, and at the boundary there is a discontinuous change in the
density of zeros. Assume the electric potentials $\phi_i(z)$, $i=1,2$, solve the Laplace
equation \eqref{eq:laplace} with the corresponding density $\rho_i(z)$. With
this a global solution is 
\begin{align}
\phi(z)=\left\{\begin{matrix}\phi_1(z)&, z\in I\\\phi_2(z)&,z\in
II\end{matrix}\right.  
\end{align} 
and we demand continuity of $\phi(z)$ at
the boundary. Let us now focus on the behavior of $\phi(z)$ at the
intersection of the boundary with the real time axis. If we choose the
$x$-$y$-coordinate frame as indicated in Fig. \ref{fig:density_area}, the
continuity of $\phi(z)$ implies that on the boundary 
\begin{align}
\frac{\partial^2}{\partial y^2}\phi_1(z)-\frac{\partial^2}{\partial
y^2}\phi_2(z)=0\ .  
\end{align} 
Since we are interested in the behavior of
$\phi(z)$ in the real time axis, we transform to $t$-$y'$-coordinates,
\begin{align} 
t&=\frac{x}{\cos\alpha}+\frac{y}{\sin\alpha}\ ,\nonumber\\ 
y'&=y\ .
\end{align} 
With the Laplace equation \eqref{eq:laplace} this yields
\begin{align} 
(\cos\alpha)^{-2}\frac{\partial^2}{\partial t^2}\phi_i(z)
+(1+\sin\alpha^{-1})^2\frac{\partial^2}{\partial y'^2}\phi_i(z) =
-2\pi\rho_i(z)\ , 
\end{align} 
and, consequently, 
\begin{align}
\frac{\partial^2}{\partial t^2}\left[\phi_1(z)-\phi_2(z)\right]=
-2\pi(\cos\alpha)^2(\rho_1(z)-\rho_2(z))\ .  
\end{align} 
This means, if an area of zeros of the partition function
overlaps the real time axis, the second derivative of the free energy is
discontinuous.

\subsection{2d BCS-type models} 
In the Kitaev honeycomb model (and general
BCS-type models) the partition function is given by
\eqref{eq:bcs_dyn_part_func}, i.e., the zeros in the complex plane are given by
\begin{align} 
z_n(\vec k)=\frac{1}{2E_{\vec k}}\left[\ln\left(B_{\vec
k}^2\right)+i\pi(2n+1)\right]\ ,\ \ n\in\mathbb Z\ .\label{eq:zeros}
\end{align} 
At this point it becomes obvious, that in the thermodynamic limit
the double product over $k_x$ and $k_y$ in \eqref{eq:bcs_dyn_part_func} leads
to dense areas of zeros in the complex plane, since, generally,
$\partial_{k_x}z_n(\vec k)\neq\pm\partial_{k_y}z_n(\vec k)$. These areas of zeros
cover parts of the real time axis ($z=it$) if $Re(z_n(\vec k))=0$, i.e., if 
\begin{align}
\exists\vec q\in K\text{ such that }B_{\vec q}^2=1\ . \label{eq:cond}
\end{align} 
Dubbing the $B_{\vec q}^2=1$ isoline $\mathcal B_1\subset K$, there are intervals 
\begin{align}
T_n^*=\frac{(2n+1)\pi}{2E_{\mathcal B_1}}\ ,\quad n\in\mathbb Z
\label{eq:non-analytic-periods}
\end{align} 
on the real time axis, which are covered by areas of zeros. If the
spectrum is gapped, the beginnings $t_{n}^b$ and end points $t_n^e$ of two
consecutive intervals $T_n^*$ and $T_{n+1}^*$ are equidistant with
$t_{n+1}^b-t_n^b=\frac{\pi}{2E_{\vec q_b}}$ and
$t_{n+1}^e-t_n^e=\frac{\pi}{2E_{\vec q_e}}$, where $\vec q_{b/e}$ are momenta
minimizing/maximizing $E_{\vec k}$ on the domain given by $|B_{\vec k}|=1$. The
length of the single intervals $T_n^*$ increases linearly with $n$. However, if
the spectrum is gapless, all those intervals extend to infinity.

The condition \eqref{eq:cond} allows for a physical
interpretation; namely, the occurrence of DQPTs is through a 
continuity argument related to non-thermal mode occupation.\cite{heyl2013} 
In BCS-type models the mode occupation is given by 
\begin{align} 
\langle n_{\vec k}\rangle \equiv\langle a_{\vec k}^\dagger
a_{\vec k}\rangle=\sin^2\left(\arctan B_{\vec k}\right)\ .  
\end{align} 
This
means for all modes $\vec q$, where the condition \eqref{eq:cond} is satisfied,
the mode occupation is $\langle n_{\vec q}\rangle=1/2$. Let us assume that for
any two points in $K$ there exists a path connecting both points, along which
$\langle n_{\vec k}\rangle$ is continuous,
\footnote{Assuming conventional continuity is too strong in this case, since in 
the Kitaev model the mode occupation number is not necessarily 
continuous after quenching (see Fig. \ref{fig:occnums}).} 
and the existence of modes with $\langle n_{\vec k}\rangle< 1/2$. Both assumptions
should be true for physically relevant models and were found to hold in all cases
considered in the Kitaev model. Then, we can set up the following chain of
consequences: through the continuity condition, the existence of non-thermally 
occupied modes $\vec k^+$ with $\langle
n_{\vec k^+}\rangle\geq1/2$ implies the existence of modes $\vec q$ with 
$\langle n_{\vec q}\rangle=1/2$. This in turn is equivalent to the fulfilling of the
condition \eqref{eq:cond}, which implies the occurrence of DQPTs in the time
evolution. The mode occupation $\langle n_{\vec k^+}\rangle\geq1/2$ 
is non-thermal in the sense that it cannot be realized by Fermi-Dirac 
statistics with positive temperature.

An equivalent formulation of condition \eqref{eq:cond} is 
\begin{align}
\exists\vec q\in K\text{ such that }\Delta_{\vec q}(\alpha_0)\Delta_{\vec
q}(\alpha_1)+\epsilon_{\vec q}(\alpha_0)\epsilon_{\vec q}(\alpha_1)=0\ ,
\end{align} 
where $\alpha$ is the quench parameter of the BCS-type
Hamiltonian, (see appendix \ref{app:condition}). From this it becomes clear,
that after quenching to a gapless phase there are zeros of the dynamical
partition function on the real time axis, since $E_{\vec
q}(\alpha)=0\Leftrightarrow \epsilon_{\vec q}(\alpha)=\Delta_{\vec
q}(\alpha)=0$. In the mode occupation picture this can be interpreted as follows:
when quenching to a gapless phase, excitations cost no energy; thus, any
quench produces inverted mode occupation.

As discussed in the previous section, areas of zeros covering the real time
axis result in jumps in the second time derivative of the dynamical free energy
density if there is a jump in the density of zeros. Eq. \eqref{eq:zeros} gives
a \enquote{layer} of zeros for every $n\in\mathbb Z$. Therefore, our total density
of zeros is a sum of the densities of the individual \enquote{layers},
$\rho_\text{z}(z)=\sum_n\rho_\text{z}^n(z)$. The single layer densities are
given the Jacobi determinant of the change of variables $\vec k\to z_n(\vec
k)$,\cite{saarloos_kurtze_1984}
\begin{widetext} 
\begin{align}
&\rho_\text{z}^n(z)=\frac{1}{\pi^2}\left|\begin{matrix}\frac{\partial
\text{Re}(z_n)}{\partial k_x} & \frac{\partial \text{Re}(z_n)}{\partial
k_y}\\\frac{\partial \text{Im}(z_n)}{\partial k_x}&\frac{\partial
\text{Im}(z_n)}{\partial k_y}\end{matrix}\right|^{-1}\nonumber\\
&=\frac{1}{\pi^2}\left[\left(\frac{\partial_{k_x}B_{\vec k}^2}{2E_{\vec
k}B_{\vec k}^2}-\frac{\ln(B_{\vec k}^2)}{2E_{\vec k}^2}\partial_{k_x}E_{\vec
k}\right)\left(-\frac{(2n+1)\pi}{2E_{\vec k}^2}\partial_{k_y}E_{\vec
k}\right)-\left(\frac{\partial_{k_y}B_{\vec k}^2}{2E_{\vec k}B_{\vec
k}^2}-\frac{\ln(B_{\vec k}^2)}{2E_{\vec k}^2}\partial_{k_y}E_{\vec
k}\right)\left(-\frac{(2n+1)\pi}{2E_{\vec k}^2}\partial_{k_x}E_{\vec
k}\right)\right]^{-1}\nonumber\\ 
&=\frac{4E_{\vec k}^3B_{\vec
k}^2}{(2n+1)\pi^3}\left(\partial_{k_x}E_{\vec k}\partial_{k_y}B_{\vec
k}^2-\partial_{k_y}E_{\vec k}\partial_{k_x}B_{\vec
k}^2\right)^{-1}=\frac{4E_{\vec k}^3B_{\vec
k}^2}{(2n+1)\pi^3}\left|\begin{matrix}\frac{\partial E_{\vec k}}{\partial k_x}
& \frac{\partial E_{\vec k}}{\partial k_y}\\
\frac{\partial B_{\vec
k}^2}{\partial k_x}&\frac{\partial B_{\vec k}^2}{\partial
k_y}\end{matrix}\right|^{-1}\ ,\ \ \vec k\equiv\vec k(z_n) 
\end{align}
\end{widetext} 
At this point a more technical view of the zeros of the
partition function is useful: the zeros $z_n(\vec k)$ correspond to
intersections of the isolines 
\begin{align} 
B_{\vec
k}^2&=\exp\left(\frac{(2n+1)\pi\text{Re}(z_n(\vec k))}{\text{Im}(z_n(\vec
k))}\right)\ ,\nonumber\\ 
E_{\vec k}&=\frac{(2n+1)\pi}{2\text{Im}(z_n(\vec k))}
\end{align} 
in the momentum plane. This means that the density of zeros
$\rho_\text{z}^n(z)$ diverges at the boundary, since there $\vec\nabla E_{\vec
k}\parallel\vec\nabla B_{\vec k}^2$. Thus, when approaching the boundary of an
interval $T_n^*$ from the inside of the interval, the second time derivative of
$\text{Re}\left[f(t)\right]$ will diverge.

\section{Dynamical phase transitions in the Kitaev honeycomb
model\label{sec:kitaev_dqpt}} 
\subsection{Zeros of the dynamical partition
function in the Kitaev model} 
In the Kitaev model not only quenches to the
massless phase create inverted mode occupation. Also quenches across phase
boundaries with final parameter $\vec J_1$ in a massive phase induce critical
points in the real time evolution. It is physically reasonable to assume, that
the mode occupation number $\langle n_{\vec k}\rangle$ is sufficiently well
behaved, namely, that for any two $\vec k_0$, $\vec k_1$ there exists a path
$\vec\gamma:[0,1]\to [-\pi,\pi]^2$ with $\vec\gamma(0)=\vec k_0$ and
$\vec\gamma(1)=\vec k_1$ such that $\langle n_{\vec\gamma(s)}\rangle$,
$s\in[0,1]$, is continuous. We found this to be true for all considered cases. 
Under this prerequisite, the existence of a fully occupied
mode $\vec k^+$, $\langle n_{\vec k^+}\rangle=1$, implies that $\langle n_{\vec
q}\rangle=1/2$ somewhere, because $\langle n_{\vec k=0}\rangle=0$. $\langle
n_{\vec k^+}\rangle=1$ corresponds to $|B_{\vec k^+}|=\infty$ and this happens when
\begin{align} 
0&=u_{\vec k^+}(\vec J_0)u_{\vec k^+}(\vec J_1)+v_{\vec k^+}(\vec
J_0)v_{\vec k^+}(\vec J_1)\nonumber\\ 
\land\ 0&\neq u_{\vec k^+}(\vec J_0)v_{\vec k^+}(\vec
J_1)-u_{\vec k^+}(\vec J_1)v_{\vec k^+}(\vec J_0)\ .  
\end{align} 
One possibility to fulfill this is 
\begin{align} 
\pm 1=\frac{\epsilon_{\vec k^+}(\vec
J_0)}{E_{\vec k^+}(\vec J_0)}=-\frac{\epsilon_{\vec k^+}(\vec J_1)}{E_{\vec k^+}(\vec
J_1)}\ .\label{eq:ncond} 
\end{align} 
Now, consider a quench ending in the
$x$-phase ($J_1^x\geq J_1^y+J_1^z$) and $\vec k^+=(\pi,0)$. Then $\epsilon_{\vec
k^+}(\mathbf J_1)=2(J_1^z-J_1^x+J_1^y)$ and $\epsilon_{\vec k^+}(\vec J_1)/E_{\vec
k^+}(\vec J_1)=-1$. We find that at this point both quenches, starting from
another massive phase, 
\begin{align} 
J_0^y<J_0^x+J_0^z\ \Rightarrow\
\frac{\epsilon_{\vec k^+}(\vec J_0)}{E_{\vec k^+}(\vec J_0)}=1\ , 
\end{align} 
and from the massless phase, 
\begin{align} 
J_0^x<J_0^y+J_0^z\ \Rightarrow\
\frac{\epsilon_{\vec k^+}(\vec J_0)}{E_{\vec k^+}(\vec J_0)}=1\ , 
\end{align}
lead to non-analytic behavior because \eqref{eq:ncond} is fulfilled in both
cases. The same can be shown for quenches ending in the other massive phases,
only $\vec k^+$ needs to be chosen appropriately. This shows that in the Kitaev
model occupation inversion is produced by quenches within the massless phase or
quenches crossing phase boundaries.

\begin{figure} \center
\resizebox{.95\columnwidth}{!}{\includegraphics{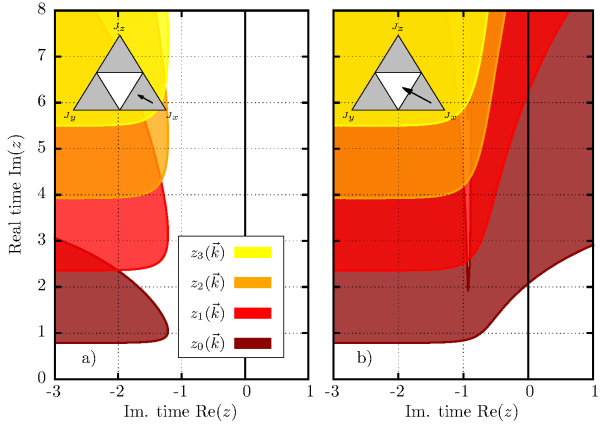}}
\caption{Distribution of zeros of the Loschmidt echo in the complex time plane
computed according to eq. \eqref{eq:zeros} for two different quenches. The
zeros form areas in the complex plane. (a) Quench within one phase. The zeros
are restricted to the left half plane and no DQPTs occur. (b) Quench to the
massless phase. The zero areas overlap the real time axis (i.e., imaginary $z$
axis) and DQPTs occur at the intersections of the boundaries of the single
areas $z_n(\vec k)$ with the real time axis. Time is measured in units of
$\sum_{\alpha} J_1^\alpha $.} \label{fig:cplane_zeros} 
\end{figure}

Figure \ref{fig:cplane_zeros} displays locations of the zeros of the Loschmidt
echo in the complex plane given by eq. \eqref{eq:zeros} for two quenches, one
within the $A_x$ phase and one from the $A_x$ phase to the massless phase. Both
panels include a phase diagram with an arrow indicating the quench parameters
$\vec J_0\to\vec J_1$. The numerical values for the parameters for this figure
and all following figures are listed in Tab. \ref{tab:params} in the
appendix. The zeros do indeed form areas, which are restricted to the left
half-plane for the quench within the massive phase but cover parts of the real
time (imaginary $z$) axis when $\vec J_0$ and $\vec J_1$ lie in different
phases.

\subsection{Real time evolution} 
\begin{figure} 
\center \resizebox{\columnwidth}{!}{\includegraphics{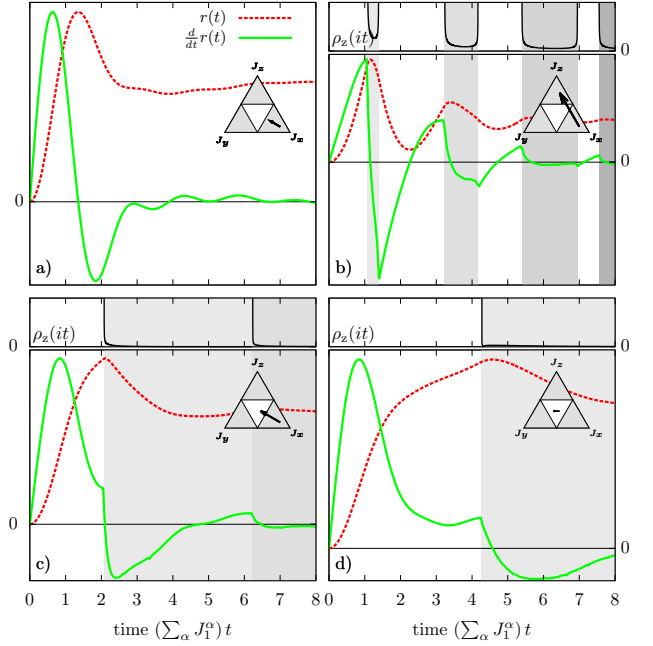}}
%\resizebox{\columnwidth}{!}{\input{figures/dqpt}}
\caption{Real time evolution of the rate function of the Loschmidt echo
\eqref{eq:rate_function_explicit_e} and its time derivative
\eqref{eq:rate_function_derivative} for various quenches. Both were obtained by
numerical evaluation of the integrals. The gray-shaded areas indicate sections
of the real time axis that are covered by areas of vanishing Loschmidt echo
(cf. Fig. \ref{fig:cplane_zeros}). If zeros of the Loschmidt echo cover parts
of the time axis, also the density of zeros on the time axis,
$\rho_\text{z}(it)$, is included. Kinks in the time derivative of the rate
function are observed when quenching across a phase boundary or within the
massless phase.} \label{fig:time_evol_2d} 
\end{figure}

On the real time axis the rate function of the
Loschmidt echo $\mathcal L(t)=|Z(it)|^2$ reads 
\begin{align}
r(t)=-\frac{1}{2\pi^2}\int_0^\pi\int_{-\pi}^\pi \D k_x \D
k_y\ln\left(\frac{\sqrt{1+2B_{\vec k}^2\cos(2E_{\vec k}t)+B_{\vec
k}^4}}{1+B_{\vec k}^2}\right)\label{eq:rate_function_explicit_e} 
\end{align}
and the time derivative is 
\begin{align} 
\dot r(t)=\frac{1}{\pi^2}\int_0^\pi\int_{-\pi}^\pi \D k_x\D k_y\frac{B_{\vec
k}^2E_{\vec k}\sin(2E_{\vec k}t)}{1+2B_{\vec k}^2\cos(2E_{\vec k}t)+B_{\vec
k}^4}\ .\label{eq:rate_function_derivative} 
\end{align} 
Figure
\ref{fig:time_evol_2d} shows the time evolution of the rate function and its
time derivative for various quenches obtained by numerical evaluation of the
corresponding integrals. The gray-shaded areas in the plots indicate the
intervals $T_n^*$ (cf. eq. \eqref{eq:non-analytic-periods}) of vanishing
partition function. The results exhibit the properties expected from the
previous considerations. The rate function is smooth for quenches within the
gapped phases; however, nonanalyticities occur when phase boundaries are
crossed in a quench or after a quench within the gapless phase.  The beginning
and end points of the critical intervals are equidistant, respectively, and for
quenches ending in the massless phase the intervals extend to infinity. As
anticipated, nonanalyticities only show up at the boundaries of the critical
intervals. Moreover, the nonanalyticities emerge as discontinuities of $\ddot
r(t)$, i.e., kinks in $\dot r(t)$.

Note that the two plots in panels a) and c) show the time evolution of the rate
function after the two quenches for which Fig. \ref{fig:cplane_zeros} shows the
location of the zeros of the partition function.

\subsection{The 1d limit} 
\begin{figure} \center
\resizebox{\columnwidth}{!}{\includegraphics{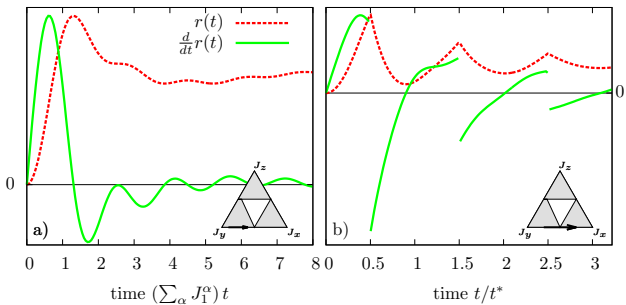}}
\caption{Time evolution of the rate function $r(t)$ of the Loschmidt echo and
its time-derivative for two quenches in the effectively one-dimensional Kitaev
model with $J_z=0$. The quench in (a) does not cross the phase boundary and the
rate function is analytic. In (b) the phase boundary is crossed and
discontinuities of $\dot r(t)$ occur at equidistant instances in time $t_n^*$.}
\label{fig:dqpt_1d} 
\end{figure}

In the limit $J_\alpha\to0$ for any
$\alpha\in\{x,y,z\}$ the 2d Kitaev model \eqref{eq:kitaev_hamiltonian} 
degenerates and becomes a set of
separate 1d spin chains. Let us consider the case $J_z=0$. The vanishing of one
of the other two parameters will give the same result due to the threefold
symmetry. For $J_z=0$ the condition for nonanalyticities \eqref{eq:cond} is
fulfilled at $\vec q$ with 
\begin{align}
\cos(q_x-q_y)=\frac{J_0^xJ_1^x+J_0^yJ_1^y}{J_0^xJ_1^y+J_1^xJ_0^y}\ .
\end{align} 
Along this line also the spectrum is constant, 
\begin{align} 
E_{\vec q}&=\sqrt{{J_1^x}^2+{J_1^y}^2+2J_1^xJ_1^y
\frac{J_0^xJ_1^x+J_0^yJ_1^y}{J_0^xJ_1^y+J_1^xJ_0^y}}\ .  
\end{align} 
Thus, the critical intervals $T_n^*$ defined in eq.
\eqref{eq:non-analytic-periods} become critical points 
\begin{align}
t_n^*=\frac{(2n+1)\pi}{2E_{\vec q}}\equiv\frac{2n+1}{2}t^*
\label{eq:crit_times_1d} 
\end{align} 
on the real time axis. 
Figure \ref{fig:dqpt_1d} shows the rate functions for two different
quenches with $J_0^z=J_1^z=0$. The quench in Fig. \ref{fig:dqpt_1d}a does not
cross a phase boundary and therefore the rate function is analytic. However, in
Fig. \ref{fig:dqpt_1d}b $\vec J_0=\left(\frac14,\frac34,0\right)$ and $\vec
J_1=\left(\frac34,\frac14,0\right)$ lie in different phases, and according to
eq. \eqref{eq:crit_times_1d} there are critical times
$t_n^*=\frac{2n+1}{2}\sqrt{\frac{5}{8}}\pi$ at which the rate function becomes
singular. In particular, these singularities are discontinuities in the first
time derivative of the rate function, not in the second as in the genuinely
two-dimensional cases. This observation underlines the fact that the continuity
of the first derivative is inherent to the higher dimensionality of the
non-degenerate Kitaev model.

\subsection{The long time limit}

In a recent work\cite{heyl_vojta_2013} it was stated that if the Loschmidt
echo $\mathcal L(t)$ supports analytic continuation, then the long real time
limit $\lim_{t\to\infty}\mathcal L(t)$ and the long imaginary time limit
$\lim_{\tau\to\infty}\mathcal L(-i\tau)$ coincide. In the context of dynamical
phase transitions this gives rise to the conjecture, that the occurrence of 
DQPTs is closely related to the long time behaviour of $\mathcal L(t)$. 
Dynamical quantum phase
transitions occur if the zeros of the dynamical partition function cross the
real time axis. This means, $r(z)$, the rate function of the Loschmidt echo, 
is non-analytic in the $\tau>0$ half plane; therefore, 
the long imaginary-time limit and the long real time
limit do not necessarily have to coincide.

On the imaginary time axis $z=\tau$, with the eigenbasis of the quenched
Hamiltonian $|\phi_n\rangle$, corresponding energies $E_n$, and
$c_n=\langle\phi_n|\psi_i\rangle$, 
\begin{align} 
\mathcal L(-i\tau)
&=\left|\langle\psi_i|e^{-H\tau}|\psi_i\rangle\right|^2
=\left|\sum_{n,n'}c_n^*c_{n'}e^{-E_{n'}\tau}\langle
\phi_n|\phi_{n'}\rangle\right|^2\nonumber\\ 
&=\left|\sum_n|c_n|^2e^{-E_n\tau}\right|^2\ .  
\end{align} 
Now, shifting the energy such that $E_0=0$ and assuming the
ground state $|\phi_0\rangle$ to be non-degenerate, 
\begin{align}
\lim_{\tau\to\infty}\mathcal L(-i\tau)=\left|c_n\right|^4
=\left|\langle\phi_0|\psi_i\rangle\right|^4=\mathcal F^4\ , \label{eq:lit_loschmidt}
\end{align} 
where $\mathcal F\equiv|\langle\phi_0|\psi_i\rangle|$ is
the fidelity. Thereby, the Loschmidt echo is connected to the fidelity in the
large imaginary time limit. 

As the Loschmidt echo is not well defined in the thermodynamic limit, one
should rather formulate eq. \eqref{eq:lit_loschmidt} in terms of the rate
function: 
\begin{align} 
\lim_{\tau\to\infty} r(-i\tau)&=-\lim_{\tau\to\infty}\lim_{N\to\infty}\frac{1}{N}\ln\mathcal
L(-i\tau)\nonumber\\
&=-\lim_{N\to\infty}\frac{1}{N}\ln\mathcal F^4\label{eq:lt_rf} 
\end{align}
According to the previous considerations this yields 
\begin{align} 
\lim_{t\to\infty}r(t)
%&=-\lim_{N\to\infty}\frac{1}{N}
%\ln\left(|\langle\psi_i|\phi_0\rangle|^4\right)\nonumber\\
&=-\lim_{N\to\infty}\frac{1}{N}\ln\mathcal F^4\ ,\label{eq:lt_loschmidt}
\end{align} 
if $r(z)$ is analytic in the $\tau>0$ half plane. This is
a quite remarkable result: In the long time limit the Loschmidt echo approaches
a value given solely by the overlap of the initial state with the ground state
of the post-quench Hamiltonian, although the stationary state will surely never
be that ground state. Quenching inevitably produces an excited state. Moreover,
some information about the initial state is preserved for all times.

\begin{figure} 
\center 
\includegraphics[width=\columnwidth]{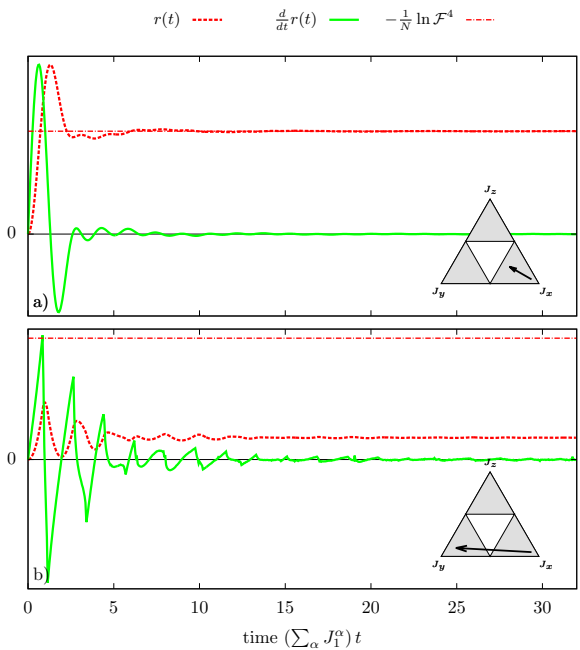}
\caption{Long-time behavior of
the rate function (a) after a quench within the $A_x$ phase and (b) after a
quench crossing phase boundaries. The rate function converges in both cases.
For the quench within the massive phase it indeed approaches the value
predicted by Heyl's and Vojta's conjecture \eqref{eq:lt_rf}, and in the other
case the limit lies well off that value.} \label{fig:lt} 
\end{figure}

For the Kitaev model the fidelity is 
\begin{align} 
\mathcal F&=|\langle\psi_i|\phi_0\rangle|
=\frac{\left|\langle0|\prod'_{\vec k}\left(1+B_{\vec k}a_{-\vec k}a_{\vec
k}\right)|0\rangle\right|}{\sqrt{\langle\psi_0|\psi_0\rangle}}\nonumber\\
&=\frac{1}{\sqrt{\langle\psi_0|\psi_0\rangle}}\overset{\text{\eqref{eq:norm}}}{=}
\exp\left(-\frac{N}{2}\int\frac{\D^2k}{4\pi^2}\ln\left(1+B_{\vec k}^2\right)\right)\ .  
\end{align} 
Thus, if above conjecture is valid, we should find 
\begin{align} 
\lim_{t\to\infty} r(t)&=\frac{1}{2\pi^2}\int \D^2k\ln\left(1+B_{\vec k}^2\right)
\label{eq:fidelity} 
\end{align} 
for quenches
within the massive phases. Figure \ref{fig:lt} shows the long time behavior of
the rate function for a quench within the $A_x$ phase and for a quench crossing
phase boundaries; indeed, the rate function converges to the value given
by the fidelity after the quench within the massive phase. In the other case,
however, the rate function seems to converge, but the value it approaches
differs from the one given by the fidelity. Various other cases were checked
and the behavior was always consistent with above mentioned conjecture.

One can explain the convergence of the rate function heuristically based on the
specific form given in eq. \eqref{eq:rate_function_explicit_e}. The expressions
for the long time limit of the rate function in eq. \eqref{eq:fidelity} and the
definition of the rate function in eq. \eqref{eq:rate_function_explicit_e} only
differ in the nominators in the argument of the logarithm, which are $1$ and
$1+B_{\vec k}^2e^{-2iE_{\vec k}t}$, respectively. In the long-time limit the
factor $e^{-2iE_{\vec k}t}$ oscillates extremely fast as a function of $\vec
k$. If $B_{\vec k}^2$ is slowly changing compared to these oscillations and
also small such that 
\begin{align} 
\ln\left(1+B_{\vec k}^2e^{-2iE_{\vec k}t}\right)
\approx B_{\vec k}^2e^{-2iE_{\vec k}t}\ , 
\end{align} 
then the
contributions of neighboring points in the momentum plane will cancel in the
integral and therefore both integrals eq. \eqref{eq:rate_function_explicit_e}
and eq. \eqref{eq:fidelity} become equal. However, if the integrand is singular,
there are areas where $|B_{\vec k}|\approx 1$ and therefore the contributions
of close-by points do not necessarily cancel. As a result the values of the
integrals differ. 

In the absence of DQPTs Eq. \eqref{eq:lt_loschmidt} can also 
be derived rigorously for
BCS-type models by considering the Taylor
expansion of the logarithm in the integrand of Eq. 
\eqref{eq:rate_function_explicit_e} and computing the time averages
of the single contributions in the power series.

\subsection{Ramping} 
It is known that dynamical quantum phase transitions also
occur if the Hamiltonian parameter is continuously ramped across a critical
point instead of quenching it.\cite{heyl2013} However, it is not clear what
happens after ramping the parameter within a gapless phase. In a gapless phase
the adiabatic theorem does not apply and it is known that also slow ramping can
produce a non-zero defect density.\cite{sengupta2008} But are these
excitations sufficient to induce dynamical quantum phase transitions?

Assume the parameter $\vec J(t)$ of the Hamiltonian is not quenched immediately
from $\vec J(t<0)= J_0$ to $\vec J(t\geq0)=\vec J_1$, but continuously
according to some protocol with $\vec J(t<0)=\vec J_0$ and $\vec J(t>t_r)=\vec
J_1$. In this case, the state of the system must at any time still be of the
form given in eq. \eqref{eq:initial_state}, because at any time $H(\vec J(t))$
can only excite both modes with opposite momenta $\vec k$, $-\vec k$ in one
$\vec k$-sector. For $t>t_r$ the dynamics are the same as after a quench,
however, the $B_{\vec k}$ will depend on the details of the ramping protocol.
As discussed above $B_{\vec k}$ is directly related to the mode occupation
number $\langle n_{\vec k}\rangle$. Thus, in order to determine, whether DQPTs
occur after ramping from $\vec J_0$ to $\vec J_1$ instead of quenching, it is
sufficient to compute the mode occupation at $t=t_r$.

In order to get the mode occupation $\langle n_{\vec k}\rangle$ we make use of
the fact that the total time evolution is simply made up by the time evolution
of independent two-level systems in the single $\vec k$-sectors and the
corresponding Hamiltonians are 
\begin{align} 
H_{\vec k}(\vec J(t))=
\frac{1}{2}\left(\begin{matrix}\epsilon_{\vec k}(\vec J(t))& \Delta_{\vec
k}(\vec J(t))\\\Delta_{\vec k}(\vec J(t))&\epsilon_{\vec k}(\vec
J(t))\end{matrix}\right)\ .  
\end{align} 
In these terms the initial state is
the ground state of $H_{\vec k}(\vec J_0)$, $H_{\vec k}(\vec J_0)|\psi_{i,\vec
k}\rangle = -E_{\vec k}(\vec J_0)|\psi_{i,\vec k}\rangle$, and the time evolved
state $|\psi_{\vec k}(t)\rangle$ can be obtained by numerical integration of
the Schr\"odinger equation. The mode occupation number after the ramping is
then given by the overlap 
\begin{align} 
\langle n_{\vec k}\rangle = |\langle \psi_{\vec k}^+|\psi_{\vec k}(t_r)\rangle|^2\ , 
\end{align} 
where $H_{\vec k}(\vec J_1)|\psi_{\vec k}^+\rangle 
= E_{\vec k}|\psi_{\vec k}^+\rangle$.

\begin{figure} 
\center
\includegraphics[width=\columnwidth]{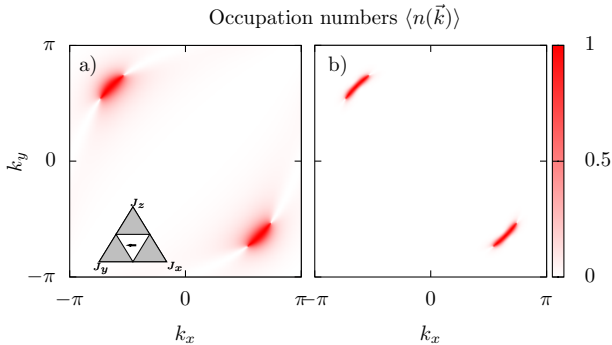} 
\caption{(a) Mode occupation $\langle n_{\vec
k}\rangle$ after quenching from $\vec J_0$ to $\vec J_1$ within the gapless
phase and (b) mode occupation after linearly ramping from $\vec J_0$ to $\vec
J_1$ with ramping time $t_r=50$.} \label{fig:occnums} 
\end{figure}

Figure \ref{fig:occnums} shows the occupation numbers for a quench within the
gapless phase and for linear ramping with 
\begin{align} 
\vec J(t)=\left\{\begin{matrix*}[l]\vec J_0&,t<0\\ \vec J_0+(\vec J_1-\vec
J_0)t/t_r&,0\leq t\leq t_r\\\vec J_1&,t>t_r\end{matrix*}\right.  
\end{align}
and ramping period $t_r=50$. As expected from the previous considerations the
quench produces regions of non-thermally occupied modes in the Brillouin-zone.
Such areas are also present after the ramping. This means, that also in the
time evolution for times $t>t_r$ there will be dynamical quantum phase
transitions.  

\subsection{Quenching the magnetic field\label{subsec:magnfield}} 
In the
presence of a magnetic field the phase $B$ (cf. Fig. \ref{fig:phase_diagram})
becomes gapped and at $J_x=J_y=J_z\equiv J$ there exists a diagonal form of the
Hamiltonian\cite{kitaev2006} with spectrum $E_{\vec k}(J,\kappa)$ as given in
eq. \eqref{eq:magn_spec}, that maps to the general two-band form (see appendix
\ref{app:bogoliubov} for details)
\begin{align} 
H(J,\kappa)=\sum_{\vec k} \vec \gamma_{\vec k}^\dagger\left(\vec b_{\vec k}(\alpha)
\cdot\vec \sigma\right)\vec \gamma_{\vec k}\ , 
\end{align} 
where
$\vec\sigma=(\sigma_x,\sigma_y,\sigma_z)^T$ is the vector of Pauli matrices,
$\gamma_{\vec k}^\dagger=(d_{\vec k}^\dagger,d_{-\vec k})$, and 
\begin{align}
b_{\vec k}(J,\kappa)=\frac{1}{2}\left(\begin{matrix}\text{Re}(\tilde\Delta_{\vec
k}(J,\kappa))\\\text{Im}(\tilde\Delta_{\vec k}(J,\kappa))\\\tilde\epsilon_{\vec
k}(J,\kappa)\end{matrix}\right)\ .  
\end{align} 
The magnetic field $\vec h$ is
contained in the parameter $\kappa\sim\frac{h_xh_yh_z}{J^2}$. It introduces
topological order in the $B$-phase, characterized by the Chern number
\begin{align} 
\nu(\kappa)&=\frac{1}{4\pi}\int_{-\pi}^\pi\int_{-\pi}^\pi \D
k_x\D k_y\frac{\vec b_{\vec k}\cdot(\partial_{k_x}\vec b_{\vec
k}\times\partial_{k_y}\vec b_{\vec k})}{|\vec b|^3}\nonumber\\ 
&=\text{sign}(\kappa)\ .
\end{align} 
It was demonstrated that in such systems any quench crossing the
boundary between topologically distinct phases induces dynamical quantum phase
transitions.\cite{vajna2014}

\begin{figure} 
\center
\includegraphics[width=\columnwidth]{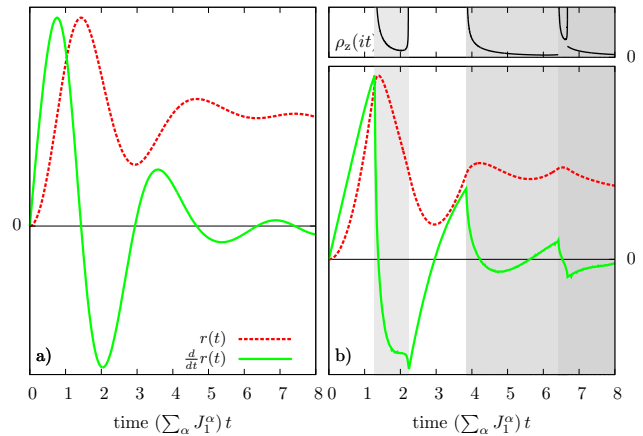} 
\caption{Time evolution of the rate function in
the Kitaev model with additional magnetic field after quenching the magnetic
field. (a) Quench within one phase, $\kappa_0=0.5\to\kappa_1=0.1$. (b) Quench
across the phase boundary, $\kappa_0=0.5\to\kappa_1=-0.1$.} \label{fig:rf_magn}
\end{figure}

Figure \ref{fig:rf_magn} shows the time evolution of the rate function after
quenching the magnetic field within one phase and between two topologically
distinct phases. As expected the signature of dynamical quantum phase
transitions shows up after the quench across the phase boundary.

\section{Conclusions} 
We demonstrated how the domains of zeros in 2d BCS-type
models differ qualitatively compared to 1d systems; namely, in the
thermodynamic limit, the zeros coalesce to dense areas of zeros in the 
complex time plane rather than lines. The
covering of intervals of the real time axis by such areas of zeros indicates
the existence of critical points in the time evolution. We showed how this
leads to dynamical quantum phase transitions as discontinuities in the second
time derivative of the dynamical free energy as opposed to discontinuities in
the first derivative known from 1d systems.

It was found that in the Kitaev honeycomb model dynamical quantum phase
transitions occur after quenches across the phase boundaries and after quenches
within the gapless phase. It is to our knowledge the first time that DQPTs are found
in a two-dimensional model after quenching without crossing an equilibrium
phase boundary. In accordance with the general considerations
regarding the dynamics of BCS-type systems, DQPTs in the Kitaev model show up
at the boundaries of the intervals $T_n^*$ on the time axis which are included in
a domain of zeros of the partition function. At these points the DQPTs appear as 
kinks in the first time derivative of the rate function of the Loschmidt
echo. As was shown to hold for any BCS-type model, the
curvature of the free energy density diverges when the boundary of such an
interval is approached from the inward. 

In degenerate versions of the Kitaev model, which effectively constitute
one-dimensional spin chains, already the first derivative of the rate function
becomes discontinuous after quenching across a phase boundary as known from
other 1d models.\cite{heyl2013} This underlines the fact that the continuity of
the first derivative is inherent to the higher dimensionality of the
non-degenerate Kitaev model. 

Moreover, we found for the Kitaev model that, in accordance with a conjecture
concerning that matter,\cite{heyl_vojta_2013} the long time stationary
state of the rate function of the Loschmidt echo
has a close connection to the occurrence of DQPTs:
if no DQPTs occur, i.e., if the rate function is analytic
in at least a half of the complex plane, the rate function approaches a value
given by the fidelity. If the rate
function is, however, non-analytic, it does in general not converge to this
value. The fact that the long time limit in absence of DQPTs is given by the fidelity
deserves particular notice, since the fidelity is the
overlap of the initial state with the ground state of the quenched Hamiltonian,
but the approached stationary state is surely an excited state. 

The examination of the mode occupation numbers after ramping the parameter of
the Hamiltonian instead of quenching it implies that the subsequent time
evolution also exhibits dynamical quantum phase transitions. Moreover, it was
demonstrated that in the presence of a magnetic field quenches between the
topologically ordered phases induce dynamical quantum phase transitions, which
was previously proven to be a general feature in the dynamics of topologically
ordered two-band models in Ref. \onlinecite{vajna2014}.

DQPTs were to date not observed in experiments. The Loschmidt echo $\mathcal
L(t)$, which shows the non-analytic behavior, is not directly connected to a
quantum mechanical observable. As mentioned above, the work density was
suggested as an measurable quantity, which could show the signature of DQPTs.
However, the Loschmidt echo becomes exponentially small with increasing system
size, whereas, strictly speaking, DQPTs occur only in the thermodynamic limit.
Experimentally measuring the signature of DQPTs in work densities will therefore
be very challenging. Experimental consequences of DQPTs in other quantities
than the work distribution function are currently being investigated. It was
shown that DQPTs are connected to timescales of the order parameter dynamics in
symmetry broken systems;\cite{heyl2013} in particular the occurrence of
DQPTs is directly related to the sudden transition from monotonous to
oscillatory decay of the order parameter after a quench.\cite{heylnov2014} 
%The 
%order parameter is observable in experiment, but it will not reveal the
%critical behaviour, which would be required for the analysis of possible
%universality.  
Alternatively, the previously mentioned generalized expectation
values could serve as measurable quantities.\cite{canovi2014}
Moreover, it was shown in a recent work that DQPTs in Ising spin models exhibit
scaling and universality and numerical results indicate that signatures of the
DQPTs can be found in the dynamics of spin correlations as power-law 
scaling, which is solely determined by the universality 
class.\cite{heyl_universality_2015} This seems to provide a very promising
opportunity for measuring DQPTs, since these quantities are accessible with
current experimental techniques.

The finding of DQPTs as discontinuities in higher order derivatives in higher
dimensional systems raises the question of a classification of dynamical
quantum phase transitions. Canovi et al. \cite{canovi2014} suggested a
formalism for such a classification. They related discontinuities in
generalized expectation values and coexisting solutions to a first order
transition. Here, we found a discontinuity in the second derivative of the
dynamical free energy density. In future work it should be investigated 
how the findings of discontinuities in higher derivatives of the
dynamical free energy density tie in with their definition.

\begin{acknowledgments}
The authors thank N. Abeling, M. Heyl and, B. Blobel 
for valuable discussions.
S.K. acknowledges
support through SFB Grant No. 1073 (project B03) 
of the Deutsche Forschungsgemeinschaft (DFG).
\end{acknowledgments}

\appendix \section{Bogoliubov transformation and post-quench
eigenbasis\label{app:bogoliubov}} 
Consider the general Hamiltonian
\begin{align} 
H(\alpha)=\sum_{\vec k} \vec \gamma_{\vec k}^\dagger\left(\vec
b_{\vec k}(\alpha)\cdot\vec \sigma\right)\vec \gamma_{\vec
k}\label{eq:2bandham} 
\end{align} 
with $\vec\sigma$ the vector of Pauli matrices 
\begin{align}
\sigma_x=\left(\begin{matrix}0&1\\1&0\end{matrix}\right)\ ,\ \
\sigma_y=\left(\begin{matrix}0&-i\\i&0\end{matrix}\right)\ ,\ \
\sigma_z=\left(\begin{matrix}1&0\\0&-1\end{matrix}\right)\ , 
\end{align} 
and $\vec\gamma_{\vec k}$ containing the creation and 
annihilation operators
\begin{align} 
\gamma_{\vec k}=\left(\begin{matrix}d_{\vec k}\\ d_{-\vec
k}^\dagger\end{matrix}\right)\ .  
\end{align} 
For $\alpha=\vec J$ and $\vec
b_{\vec k}(\vec J)=(\Delta_{\vec k}(\vec J)/2, 0, \epsilon_{\vec k}(\vec J)/2)$
this gives the Hamiltonian of the Kitaev model as in eq.
\eqref{eq:jw-hamiltonian}. The unitary transformation 
\begin{align}
\left(\begin{matrix}a_{\vec k}^\alpha\\ {a_{-\vec
k}^\alpha}^\dagger\end{matrix}\right)=W(\alpha)\vec\gamma_{\vec
k}=\left(\begin{matrix}u_{\vec k}(\alpha)&v_{\vec k}(\alpha)\\-v_{\vec
k}(\alpha)^*&u_{\vec k}(\alpha)^*\end{matrix}\right)\left(\begin{matrix}d_{\vec
k}\\ d_{-\vec k}^\dagger\end{matrix}\right)\label{eq:bogoliubov-trafo}
\end{align} 
that brings the Hamiltonian \eqref{eq:2bandham} into diagonal form
is the Bogoliubov transformation. Plugging eq. \eqref{eq:bogoliubov-trafo} into
eq. \eqref{eq:2bandham} and demanding that all off-diagonal terms vanish yields
\begin{align} 
0=&d_{\vec k}^z(u_{\vec k}v_{-\vec k}-u_{\vec -k}v_{\vec k})\nonumber\\
&+(d_{\vec k}^x-id_{\vec k}^y)u_{\vec -k}u_{\vec k}+(d_{\vec k}^x+id_{\vec
k}^y)v_{\vec -k}v_{\vec k}\ , 
\end{align} 
where all the dependencies on
$\alpha$ have been dropped for the sake of brevity. Since $W$ is unitary,
$u_{\vec k}$ and $v_{\vec k}$ have the general form 
\begin{align} 
u_{\vec k}=\cos\theta_{\vec k}e^{i\phi_{\vec k}}\ ,\ \ v_{\vec k}=
\sin\theta_{\vec k}e^{i\psi_{\vec k}}\ .  
\end{align} 
By choosing 
\begin{align} 
\phi_{\vec k}=-\psi_{\vec k}
=-\frac{1}{2}\arctan\left(-\frac{d_{\vec k}^y}{d_{\vec k}^x}\right) 
\end{align} 
the equation above becomes real, 
\begin{align}
0=&d_{\vec k}^z\left(\sin\theta_{\vec k}\cos\theta_{\vec k}-\cos\theta_{-\vec
k}\sin\theta_{\vec k}\right)\nonumber\\ 
&+|d_{\vec k}^{xy}|\left(\cos\theta_{\vec
k}\cos\theta_{-\vec k}-\sin\theta_{\vec k}\sin\theta_{-\vec k}\right)\ ,
\end{align} 
where $d_{\vec k}^{xy}\equiv d_{\vec k}^x-id_{\vec k}^y$ was
introduced. Then for $a_{-\vec k}^\dagger = (a_{-\vec k})^\dagger$ to hold,
$\theta_{\vec k}$ must be an odd function of $\vec k$. Thus, 
\begin{align}
&0=-d_{\vec k}^z\sin(2\theta_{\vec k})+|d_{\vec k}^{xy}|\cos(2\theta_{\vec
k})\nonumber\\ 
\Rightarrow\ &\tan(2\theta_{\vec k})=\frac{|d_{\vec k}^{xy}|}{d_{\vec
k}^z} 
\end{align} 
and this yields 
\begin{align} 
|u_{\vec
k}|^2&=\cos^2\theta_{\vec k}=\frac{1}{2}\left(1+\frac{\epsilon_{\vec
k}}{E_{\vec k}}\right)\ ,\nonumber\\ 
|v_{\vec k}|^2&=\sin^2\theta_{\vec
k}=\frac{1}{2}\left(1-\frac{\epsilon_{\vec k}}{E_{\vec k}}\right)\ .
\end{align} 
In the end it is left to choose the signs appropriately such that
$u_{\vec k}=u_{-\vec k}$ and $v_{\vec k}=-v_{-\vec k}$, e.g.,  
\begin{align}
u_{\vec k}&=e^{i\phi_{\vec k}}\sqrt{\frac{1}{2}\left(1+\frac{\epsilon_{\vec
k}}{E_{\vec k}}\right)}\ ,\nonumber\\ 
v_{\vec k}&=\text{sign}\left(d_{\vec
k}^x\right)e^{-i\phi_{\vec k}}\sqrt{\frac{1}{2}\left(1-\frac{\epsilon_{\vec
k}}{E_{\vec k}}\right)}\ .  
\end{align} 
With this transformation the Hamiltonian \eqref{eq:2bandham} becomes 
diagonal 
\begin{align}
H(\alpha)=\sum_{\vec k} \frac{E_{\vec k}(\alpha)}{2}\left({a_{\vec
k}^\alpha}^\dagger a_{\vec k}^\alpha-a_{-\vec k}^\alpha {a_{-\vec
k}^\alpha}^\dagger\right) 
\end{align} 
with $E_{\vec k}(\alpha)=|\vec b_{\vec k}(\alpha)|$.

To compute the quench dynamics one needs the connection of the degrees of
freedom $a_{\vec k}^{\alpha_0}$ that diagonalise the initial Hamiltonian
$H(\alpha_0)$ and the degrees of freedom $a_{\vec k}^{\alpha_1}$ diagonalising
the final Hamiltonian $H(\alpha_1)$. This connection is given by two subsequent
Bogoliubov transformations, 
\begin{align} 
\left(\begin{matrix}
a_{\vec k}^{\alpha_1}\\{a_{-\vec k}^{\alpha_1}}^\dagger\end{matrix}\right)
&=W(\alpha_1)W(\alpha_0)^\dagger\left(\begin{matrix}a_{\vec
k}^{\alpha_0}\\{a_{-\vec k}^{\alpha_0}}^\dagger\end{matrix}\right)\nonumber\\
&=\left(\begin{matrix}U_{\vec k}(\alpha_0,\alpha_1) & V_{\vec
k}(\alpha_0,\alpha_1)\\-V_{\vec k}(\alpha_0,\alpha_1)^*&U_{\vec
k}(\alpha_0,\alpha_1)^*\end{matrix}\right)\left(\begin{matrix}a_{\vec
k}^{\alpha_0}\\{a_{-\vec k}^{\alpha_0}}^\dagger\end{matrix}\right) 
\end{align}
with 
\begin{align} 
U_{\vec k}(\alpha_0,\alpha_1) &= u_{\vec
k}(\alpha_1)u_{\vec k}(\alpha_0)^*+v_{\vec k}(\vec J_1)v_{\vec k}(\alpha_0)^*
\nonumber\\
V_{\vec k}(\alpha_0,\alpha_1)&= u_{\vec k}(\alpha_0)v_{\vec
k}(\alpha_1)-u_{\vec k}(\alpha_1)v_{\vec k}(\alpha_0)\ .  
\end{align} 
Since $|\psi_i\rangle$ is the ground state of $H(\vec J_0)$ 
\begin{align} 
a_{\vec
k}^{\alpha_0}|\psi_i\rangle&=\left(U_{\vec k}(\alpha_0,\alpha_1)a_{\vec
k}^{\vec J_1}-V_{\vec k}(\alpha_0,\alpha_1){a_{-\vec
k}^{\alpha_1}}^\dagger\right)|\psi_i\rangle\nonumber\\
&=0\label{eq:expansion_condition} 
\end{align} 
must hold. Moreover, the ground
state of a BCS-type Hamiltonian has vanishing total momentum; thus,
\begin{align} 
|\psi_i\rangle&=\frac{1}{\mathcal N}{\prod_{\vec
k}}'\left(1+B_{\vec k}(\alpha_0,\alpha_1){a_{\vec
k}^{\alpha_1}}^\dagger{a_{-\vec
k}^{\alpha_1}}^\dagger\right)|0;\alpha_1\rangle\nonumber\\ &=\frac{1}{\mathcal
N}\exp\left(\sum_{\vec k}B_{\vec k}(\alpha_0,\alpha_1){a_{\vec
k}^{\alpha_1}}^\dagger{a_{-\vec
k}^{\alpha_1}}^\dagger\right)|0;\alpha_1\rangle\ ,\label{eq:expansion_ansatz}
\end{align} 
where the coefficients $B_{\vec k}(\alpha_0,\alpha_1)$ are to be
determined, $\mathcal N$ is a normalization constant, and $|0;\alpha_1\rangle$
denotes the vacuum of the post-quench fermions: $a_{\vec
k}^{\alpha_1}|0;\alpha_1\rangle=0$. Plugging eq. \eqref{eq:expansion_ansatz}
into eq. \eqref{eq:expansion_condition} yields 
\begin{align} 
&\left(U_{\vec
k}(\alpha_0,\alpha_1)a_{\vec k}^{\alpha_1}-V_{\vec
k}(\alpha_0,\alpha_1){a_{-\vec k}^{\alpha_1}}^\dagger\right)
\nonumber\\
&\hspace{1cm}\times{\prod_{\vec k'}}'\left(1+B_{\vec
k'}(\alpha_0,\alpha_1){a_{\vec k'}^{\alpha_1}}^\dagger{a_{-\vec
k'}^{\alpha_1}}^\dagger\right)|0\rangle_{\alpha_1}\nonumber\\ 
&=\left(U_{\vec
k}(\alpha_0,\alpha_1)B_{\vec k}(\alpha_0,\alpha_1)-V_{\vec
k}(\alpha_0,\alpha_1)\right){a_{-\vec k}^{\alpha_1}}^\dagger
\nonumber\\
&\hspace{1cm}\times{\prod_{\vec k'\neq\vec k}}'\left(1+B_{\vec
k'}(\alpha_0,\alpha_1){a_{\vec k'}^{\alpha_1}}^\dagger{a_{-\vec
k'}^{\alpha_1}}^\dagger\right)|0;\alpha_1\rangle\nonumber\\ 
&=0\ , 
\end{align} 
which holds for 
\begin{align} 
B_{\vec k}(\alpha_0,\alpha_1)&=\frac{V_{\vec
k}(\alpha_0,\alpha_1)}{U_{\vec k}(\alpha_0,\alpha_1)}\nonumber\\ 
&=\frac{u_{\vec
k}(\alpha_0)v_{\vec k}(\alpha_1)-u_{\vec k}(\alpha_1)v_{\vec
k}(\alpha_0)}{u_{\vec k}(\alpha_0)u_{\vec k}(\alpha_1)+v_{\vec
k}(\alpha_0)v_{\vec k}(\alpha_1)}\ .  
\end{align}

\section{Condition for real time zeros of the partition
function\label{app:condition}} 
The condition $B_{\vec k}^2=1$ (cf. eq.
\eqref{eq:cond}) can be rearranged as follows: We have 
\begin{widetext}
\begin{align} 
&1=|B_{\vec k}|\nonumber\\ 
\Leftrightarrow\
&\left|\sqrt{1+\frac{\epsilon_{\vec k}(\vec J_0)}{E_{\vec k}(\vec
J_0)}}\sqrt{1+\frac{\epsilon_{\vec k}(\vec J_1)}{E_{\vec k}(\vec
J_1)}}+\text{sign}(\Delta_{\vec k}(\vec J_0)\Delta_{\vec k}(\vec
J_1))\sqrt{1-\frac{\epsilon_{\vec k}(\vec J_0)}{E_{\vec k}(\vec
J_0)}}\sqrt{1-\frac{\epsilon_{\vec k}(\vec J_1)}{E_{\vec k}(\vec
J_1)}}\right|\nonumber\\ 
&=\left|\text{sign}(\Delta_{\vec k}(\vec
J_1))\sqrt{1+\frac{\epsilon_{\vec k}(\vec J_0)}{E_{\vec k}(\vec
J_0)}}\sqrt{1-\frac{\epsilon_{\vec k}(\vec J_1)}{E_{\vec k}(\vec
J_1)}}-\text{sign}(\Delta_{\vec k}(\vec J_0))\sqrt{1-\frac{\epsilon_{\vec
k}(\vec J_0)}{E_{\vec k}(\vec J_0)}}\sqrt{1+\frac{\epsilon_{\vec k}(\vec
J_1)}{E_{\vec k}(\vec J_1)}}\right|\nonumber\\ 
\Leftrightarrow\
&\left(1+\frac{\epsilon_{\vec k}(\vec J_0)}{E_{\vec k}(\vec
J_0)}\right)\left(1+\frac{\epsilon_{\vec k}(\vec J_1)}{E_{\vec k}(\vec
J_1)}\right)+\left(1-\frac{\epsilon_{\vec k}(\vec J_0)}{E_{\vec k}(\vec
J_0)}\right)\left(1-\frac{\epsilon_{\vec k}(\vec J_1)}{E_{\vec k}(\vec
J_1)}\right)\nonumber\\ 
&+2\text{sign}(\Delta_{\vec k}(\vec J_0)\Delta_{\vec k}(\vec
J_1))\sqrt{\left(1-\left(\frac{\epsilon_{\vec k}(\vec J_0)}{E_{\vec k}(\vec
J_0)}\right)^2\right)\left(1-\left(\frac{\epsilon_{\vec k}(\vec J_1)}{E_{\vec
k}(\vec J_1)}\right)^2\right)}\nonumber\\ 
&=\left(1+\frac{\epsilon_{\vec k}(\vec
J_0)}{E_{\vec k}(\vec J_0)}\right)\left(1-\frac{\epsilon_{\vec k}(\vec
J_1)}{E_{\vec k}(\vec J_1)}\right)+\left(1-\frac{\epsilon_{\vec k}(\vec
J_0)}{E_{\vec k}(\vec J_0)}\right)\left(1+\frac{\epsilon_{\vec k}(\vec
J_1)}{E_{\vec k}(\vec J_1)}\right)\nonumber\\ 
&-2\text{sign}(\Delta_{\vec k}(\vec
J_0)\Delta_{\vec k}(\vec J_1))\sqrt{\left(1-\left(\frac{\epsilon_{\vec k}(\vec
J_0)}{E_{\vec k}(\vec J_0)}\right)^2\right)\left(1-\left(\frac{\epsilon_{\vec
k}(\vec J_1)}{E_{\vec k}(\vec J_1)}\right)^2\right)}\nonumber\\ 
\Leftrightarrow &\
\frac{\epsilon_{\vec k}(\vec J_0)\epsilon_{\vec k}(\vec J_1)}{E_{\vec k}(\vec
J_0)E_{\vec k}(\vec J_1)}=-\text{sign}(\Delta_{\vec k}(\vec J_0)\Delta_{\vec
k}(\vec J_1))\sqrt{\left(1-\left(\frac{\epsilon_{\vec k}(\vec J_0)}{E_{\vec
k}(\vec J_0)}\right)^2\right)\left(1-\left(\frac{\epsilon_{\vec k}(\vec
J_1)}{E_{\vec k}(\vec J_1)}\right)^2\right)}\nonumber\\ 
\Rightarrow\
&1=\left(\frac{\epsilon_{\vec k}(\vec J_0)}{E_{\vec k}(\vec
J_0)}\right)^2+\left(\frac{\epsilon_{\vec k}(\vec J_1)}{E_{\vec k}(\vec
J_1)}\right)^2 
\end{align} 
\end{widetext} 
Plugging this into the second last line we find the additional condition
\begin{align} 
\text{sign}(\Delta_{\vec k}(\vec J_0)\epsilon_{\vec k}(\vec J_0))
=-\text{sign}(\Delta_{\vec k}(\vec J_1)\epsilon_{\vec k}(\vec J_1))\ , 
\end{align} 
which allows to write
\begin{widetext} 
\begin{align} 
1=|B_{\vec k}|\ \Leftrightarrow\
\left[1=\left(\frac{\epsilon_{\vec k}(\vec J_0)}{E_{\vec k}(\vec
J_0)}\right)^2+\left(\frac{\epsilon_{\vec k}(\vec J_1)}{E_{\vec k}(\vec
J_1)}\right)^2\ \land\ -1=\frac{\text{sign}(\epsilon_{\vec k}(\vec
J_1)\Delta_{\vec k}(\vec J_1))}{\text{sign}(\epsilon_{\vec k}(\vec
J_0)\Delta_{\vec k}(\vec J_0))}\right]\ .  
\end{align} 
\end{widetext} 
Plugging
in $E_{\vec k}(J)^2=\epsilon_{\vec k}(J)^2+\Delta_{\vec k}(J)^2$ yields
\begin{align} 
1=|B_{\vec k}(J_0,J_1)|\ \Leftrightarrow\ \Delta_{\vec k}(\vec
J_0)\Delta_{\vec k}(\vec J_1)+\epsilon_{\vec k}(\vec J_0)\epsilon_{\vec k}(\vec
J_1)=0\ .\label{eq:Bcond_equiv} 
\end{align} 
So, for the emergence of a
non-analyticity at a given time $t=t^*$ we get two (simplified) conditions:
\begin{align} 
E_{\vec k}(\vec J_1)&=\frac{(2n+1)\pi}{2t^*} \equiv C_{t^*}^n\
,\label{eq:cond1}\\ 
0&=\Delta_{\vec k}(\vec J_0)\Delta_{\vec k}(\vec
J_1)+\epsilon_{\vec k}(\vec J_0)\epsilon_{\vec k}(\vec J_1)\label{eq:cond2}
\end{align}

\section{Quench/ramping parameters} \label{app:params} 
\begin{table}[!h]
\center \begin{tabular}{ccccccccc} \hline\hline Figure & $J_0^x$ & $J_0^y$ &
$J_0^z$ & $J_1^x$ & $J_1^y$ & $J_1^z$&$\kappa_0$&$\kappa_1$\\\hline
\ref{fig:cplane_zeros}a) & 0.8 & 0.1 & 0.1 & 0.6 & 0.2 & 0.2 & 0 & 0\\
\ref{fig:cplane_zeros}b) & 0.8 & 0.1 & 0.1 & 0.4  & 0.3 & 0.3 & 0 & 0\\\hline
\ref{fig:time_evol_2d}a) & 0.8 & 0.1 & 0.1 & 0.6 & 0.2 & 0.2 & 0 & 0\\
\ref{fig:time_evol_2d}b) & 0.8 & 0.1 & 0.1 & 0.2  & 0.1 & 0.7 & 0 & 0\\
\ref{fig:time_evol_2d}c) & 0.8 & 0.1 & 0.1 & 0.4 & 0.3 & 0.3 & 0 & 0\\
\ref{fig:time_evol_2d}d) & 0.4 & 0.3 & 0.3 & 0.3  & 0.4 & 0.3 & 0 & 0\\\hline
\ref{fig:dqpt_1d}a) & 0.1 & 0.9 & 0.0 & 0.4 & 0.6 & 0.0 & 0 & 0\\
\ref{fig:dqpt_1d}b) & 0.25 & 0.75 & 0.0 & 0.75  & 0.25 & 0.0 & 0 & 0\\\hline
\ref{fig:lt}a) & 0.9 & 0.05 & 0.05 & 0.6 & 0.2 & 0.2 & 0 & 0\\ \ref{fig:lt}b) &
0.9 & 0.05 & 0.05 & 0.1  & 0.8 & 0.1 & 0 & 0\\\hline \ref{fig:occnums}a) & 0.8
& 0.1 & 0.1 & 0.6 & 0.2 & 0.2 & 0 & 0\\ \ref{fig:occnums}b) & 0.8 & 0.1 & 0.1 &
0.4  & 0.3 & 0.3 & 0 & 0\\\hline \ref{fig:rf_magn}a) & 1/3 & 1/3 & 1/3 & 1/3  &
1/3 & 1/3 & 0.5 & 0.1\\ \ref{fig:rf_magn}b) & 1/3 & 1/3 & 1/3 & 1/3  & 1/3 &
1/3 & 0.5 & -0.1\\ \hline\hline
\end{tabular} 
\caption{Summary of the quench
parameters used for the figures in the main text.} \label{tab:params}
\end{table}
%\begin{acknowledgments} put your acknowledgments here.  \end{acknowledgments}

\bibliography{refs}

\end{document}